\documentclass[12pt,preprint,apj]{emulateapj}

\usepackage{epstopdf}

\shorttitle{Virgo Ultra-faint Dwarf Galaxy}
\shortauthors{Jang \& Lee}

\begin{document}

\title{
Discovery of an Ultra-faint Dwarf Galaxy in the Intracluster Field of the Virgo Center, a Fossil of the First Galaxies?}
\author{ In Sung Jang \&  Myung Gyoon Lee}
\affil{Astronomy Program, Department of Physics and Astronomy, Seoul National University, Gwanak-gu, Seoul 151-742, Korea}
\email{ isjang@astro.snu.ac.kr, mglee@astro.snu.ac.kr}


\begin{abstract}
Ultra-faint dwarf galaxies are newcomers among galaxies, and are the faintest galaxies in the observed universe.
To date, they have only been found around the Milky Way Galaxy and M31 in the Local Group. 
We present the discovery of an UFD in the intracluster field in the core of the Virgo cluster
(Virgo UFD1), which is far from any massive galaxies. 
The color-magnitude diagram of the resolved stars in this galaxy shows a narrow red giant branch, similar to those of metal-poor globular clusters in the Milky Way.
We estimate its distance by comparing the red giant branch with isochrones, and we obtain a value $16.4\pm0.4$ Mpc. 
This shows that it is indeed a member of the Virgo cluster.
From the color of the red giants we estimate its mean metallicity to be very low, [Fe/H] $= -2.4\pm0.4$.
Its absolute $V$-band magnitude and effective radius are derived to be
$M_V = -6.5\pm0.2$ and $r_{\rm eff} = 81\pm7$ pc, much fainter and smaller than the classical dwarf spheroidal galaxies.
Its central surface  brightness is estimated to be as low as $\mu_{V,0} = 26.37\pm 0.05$ mag arcsec$^{-2}$.
Its properties are similar to those of the Local Group analogs. 
No evidence of tidal features are found in this galaxy.
Considering its narrow red giant branch with no asymptotic giant branch stars, low metallicity, and location, it may be a fossil remnant of the first galaxies. 
\end{abstract}


\section{Introduction}

Ultra-faint dwarf galaxies (UFDs) are the faintest among the known galaxies in the observed universe. 
They are much fainter ($M_V \gtrsim -8$)  and smaller ($r_{\rm eff} \lesssim 300$ pc) than classical dwarf spheroidal galaxies (dSphs; see the reviews by \citet{mcc12,san12, bel13}, and references therein).  Although UFDs have very low stellar mass ($M_* <10^6 M_\odot$), they have dynamical masses comparable to those of dSphs. Thus they have exceedingly large values for mass to light ratio, $M/L \sim 10^2$ to $10^4 $ $M_\odot / L_\odot$ \citep{sim07,mcc12}, showing that they belong to the most dark matter dominated stellar systems in the universe.
Stellar populations in UFDs are old and very metal-poor ([Fe/H] $< -2$) \citep{oka12,san12,mcc12,vin14}.
Deep Hubble Space Telescope (HST) photometry shows that most stars in UFDs were formed in a single burst before 12 Gyrs ago,
which is in contrast to the star formation history of the classical dSphs that show often multiple bursts over an extended period 
\citep{bro13,wei14}.
Thus UFDs are strong candidates for the fossil remnants of the first galaxies that finished forming stars before the epoch of reionization in the local universe  \citep{ric05,ric10,bov11a,bov11b,wei14,vin14}. 
UFDs provide strong clues
for understanding 
two well-known dwarf problems in cosmology: missing satellite dwarf problem and 
satellite dwarf quenching problem \citep{moo99,bul01,kra10,bel13,whe14}.

To date UFDs were found only around the Milky Way Galaxy and M31 in the Local Group
\citep{mcc12,san12,bel14}. 
It is mainly due to the advent of the Sloan Digital Sky Survey (SDSS) that they were discovered in the local universe (e.g., \citep{wil05a,wil05b}).
Wide field surveys such as the SDSS allowed to search for these kinds of objects in the Local Group \citep{bel14,lae14}.
However, it is very difficult to find UFDs beyond the Local Group with current facilities, because of their faintness and smallness.
In this study we present  the discovery of a UFD in the Virgo cluster 
that is as distant as 16 Mpc, taking advantage of high resolution in the very deep images in the HST archive. 

\section{Data}

\begin{figure}
\centering
\includegraphics[scale=0.85]{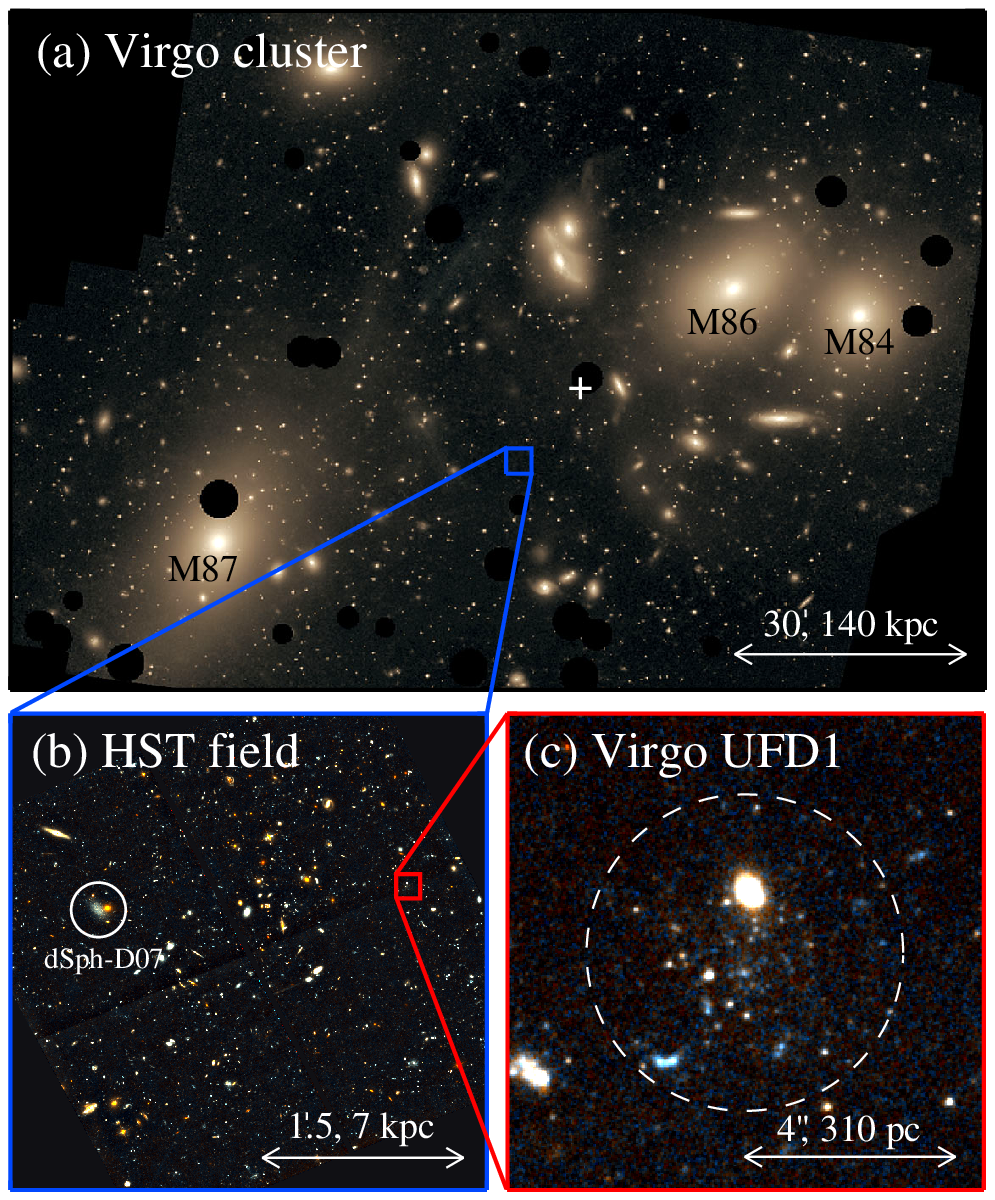} 
\caption{
(a) Identification of the HST field used in this study on a deep color image of the Virgo core by \citet{mih05}. North is up, and east to the left. 
The white cross indicates the number density center of Virgo cluster members given by \citet{bin87}.
(b) A color map of the  HST field. A dwarf spheroidal galaxy, dSph-D07, investigated by \citet{dur07} is marked by white circle, and the new Virgo UFD1 is indicated by red square. 
(c) A $10\arcsec\times10\arcsec$ section of the HST field of Virgo UFD1. 
The dashed-line circle indicates the tidal radius of Virgo UFD1.
Note that most resolved stars are red giants in the galaxy.}
\label{fig_finder}
\end{figure}

We used deep images of an isolated intracluster field close to the Virgo center  available in the archive (Prop. ID : 10131), 
as marked in Figure 1. 
Figure \ref{fig_finder}(a) shows the location of this field 
on the  very deep 
image of the Virgo Core region given by \citet{mih05}.
 This field is located far (about 180 kpc or more) from massive galaxies such as M87, M86 and M84, where the diffuse light from massive galaxies is barely recognized. It is in the region with the lowest sky surface brightness in the core of Virgo, although it is
close to the center of the Virgo cluster. 
 
These images were obtained with the F606W and F814W filters in the Advanced Camera for Surveys on board the $HST$.  
These images were used in  previous studies:
the discovery of four intracluster globular clusters  \citep{wil07a}, a study of the metallicity distribution function of the intracluster stars
\citep{wil07b}, and  the discovery of a new resolved dSph in Virgo \citep{dur07}.

We constructed deeper drizzled images for each filter by combining single-exposure images, as done in \citet{jan14}. The pixel scale in the drizzled images is
$0\farcs03$ pixel$^{-1}$, and the FWHM of point sources is $2.9 \sim 3.0$ pixels.
Total exposure times are 63440 s and 26880 s for the F606W and F814W images, respectively. 
Figure \ref{fig_finder}(b) displays the color image constructed from the drizzled F606W and F814W images.
 IRAF/DAOPHOT \citep{ste94} was applied to the drizzled images to derive the point-spread function fitting photometry of the point sources. We then calibrated and transformed them onto Johnson-Cousins $VI$ magnitudes following \citep{sir05}. 

\section{Result}

\subsection{Discovery of a New UFD}

The dSph galaxy discovered in the same images  and studied by \citet{dur07} (called Virgo dSph-D07 hereafter), is easily seen as a relatively large system of resolved stars, as shown in Figure 1 of \citet{dur07} and Figure \ref{fig_finder} in this study.
Its $F606W$ magnitude is $ V= 20.56\pm0.05$ 
($M_V = -10.6\pm0.2$), and its effective radius 
is $r_{\rm eff} =3\farcs0\pm0\farcs5$ ($260 \pm43$ pc),
which is derived from the S\'ersic profile fit, $n = 0.59 \pm 0.12$ and $r_0=3\farcs34\pm0.06$ given by \citet{dur07} \citep{gra05}.
UFDs are, if any, expected to be much fainter and smaller and have lower surface brightness than this galaxy in the same images. 
Considering this, we visually searched for small groups of faint point sources embedded in the faint diffuse background with low surface brightness.

Through inspection of the entire field, we found only one UFD candidate in the position as marked in Figure \ref{fig_finder}(b). It indeed looks much fainter and smaller than Virgo dSph-D07 in the figure. We call this new galaxy, Virgo UFD1 hereafter. 
A zoomed in image of the galaxy in Figure 1(c) shows a small grouping of stars as well as faint diffuse background emission in an almost circular area with $\approx 2\arcsec$ radius. 
One bright elongated object north of Virgo UFD1 
is considered to be a background galaxy because no stars are resolved in the galaxy.   
Virgo UFD1 is the first UFD discovered beyond the Local Group.

\subsection{Color--Magnitude Diagram of the Resolved Stars}

\begin{figure}
\centering
\includegraphics[scale=0.58]{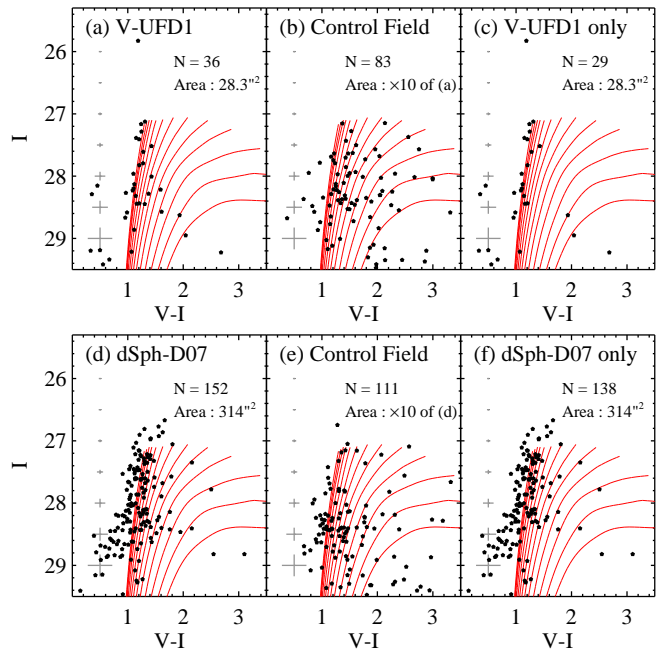} 
\caption{ $I - (V-I)$ CMDs of the resolved stars within the tidal radius of Virgo UFD1 (a) and the control field at $2.5\times r_{\rm tidal}\leq r<4.0 \times r_{\rm tidal}$ that are 10 times larger than the galaxy field (b). 
(c) The net CMD of the galaxy field obtained after statistical subtraction of the field contribution.
(d), (e), and (f) Same as (a), (b), and (c), except for Virgo dSph-D07.
Curved lines denote the 12 Gyr stellar isochrones for  [Fe/H] = $-2.4$ to $0.0$ in steps of 0.2 provided by the Dartmouth group \citep{dot08}, shifted according to
$(m-M)_0=31.08$ and $E(V-I)=0.031$.
}
\label{fig_cmd}
\end{figure}

\begin{figure*}
\centering
\includegraphics[scale=0.7]{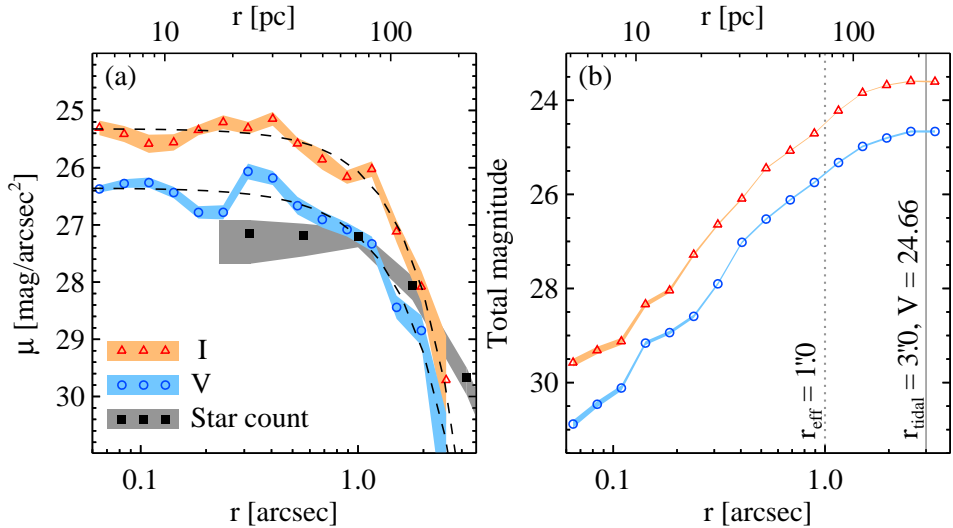} 
\caption{ 
Radial profiles of the $V$ (open circles) and $I$-band (triangles)  surface brightness (a) and integrated magnitudes (b) of Virgo UFD1 .
The bands represent the errors.
Dashed lines in (a) represent a S\'ersic law fit for each band.
The effective radius and tidal radius are marked by vertical lines in (b).
The radial profile of the density of detected stars shifted arbitrarily is also overlaid by filled squares in (a).
}
\label{fig_surf}
\end{figure*}

Figure 2(a) shows the color-magnitude diagram (CMD) of the resolved stars at $r <$ tidal radius ($r_{\rm tidal}= 3\farcs0$ 
from the center of Virgo UFD1).
We also plotted the CMD of the resolved stars in the control field at $2.5\times r_{\rm tidal} < r < 4.0\times r_{\rm tidal}$
(covering a region 10 times larger than the galaxy region to show a larger number of stars) for comparison in Figure 2(b).
We statistically subtracted the contribution of the field stars in the galaxy CMD using the control field CMD, and displayed the resulting CMD in Figure 2(c).
We also showed the CMDs of Virgo dSph-D07 as a reference to study the resolved stars in Virgo UFD1. 
We derived the photometry of the resolved stars in this galaxy 
using the same procedures as were used for Virgo UFD1. 
Figures 2(d), (e), and (f) display the CMD of the Virgo dSph-D07 region,
the CMD of the control field for this galaxy (10 times larger than the galaxy field), and the field-subtracted CMD, respectively.
We also overlayed isochrones for 12 Gyr age with [Fe/H]$=-2.4$ to 0.0 (with a step of 0.2) in the Dartmouth Group \citep{dot08}, shifted according to the distance modulus,
$(m-M)_0 = 31.08$ \citep{wil07b,dur07} and foreground reddening $E(V-I)=0.031$ \citep{sch11}.

Figure 2(c) shows that most of the stars in Virgo UFD1 have a narrow range of color,
$1.0<(V-I)<1.4$, and they appear to be located along the old red giant branch (RGB),
similar to those of the metal-poor globular clusters in the Milky Way Galaxy. 
The magnitude of the brightest star among these stars is $I\sim27.1$. 
One lonely star 1.3 mag above the brightest part of the RGB is probably a field object.
There are two stars much bluer than the RGB ($(V-I)\sim 0.4$ and $I\sim 28.2$). 
Both stars are blended with other objects. Moreover, one of them is located relatively far ($r \approx 1\farcs8 $) from the center of the galaxy. 
They may not belong to this galaxy.
Comparing the CMDs for Virgo UFD1 and Virgo dSph-D07 shows that the RGB of 
Virgo UFD1 is in general very similar to that of Virgo dSph-D07 given that have a very low metallicity [Fe/H] $= -2.3\pm0.3$ \citep{dur07}, indicating that the metallicity of Virgo UFD1 is also very metal-poor. 
It is also noted that there are no asymptotic giant branch stars in Virgo UFD1, while there are a small number of AGB candidates above the tip of the RGB in Virgo dSph-D07.
This indicates that Virgo UFD1 is very old (older than 10 Gyr).

\begin{figure*}
\centering
\includegraphics[scale=0.7]{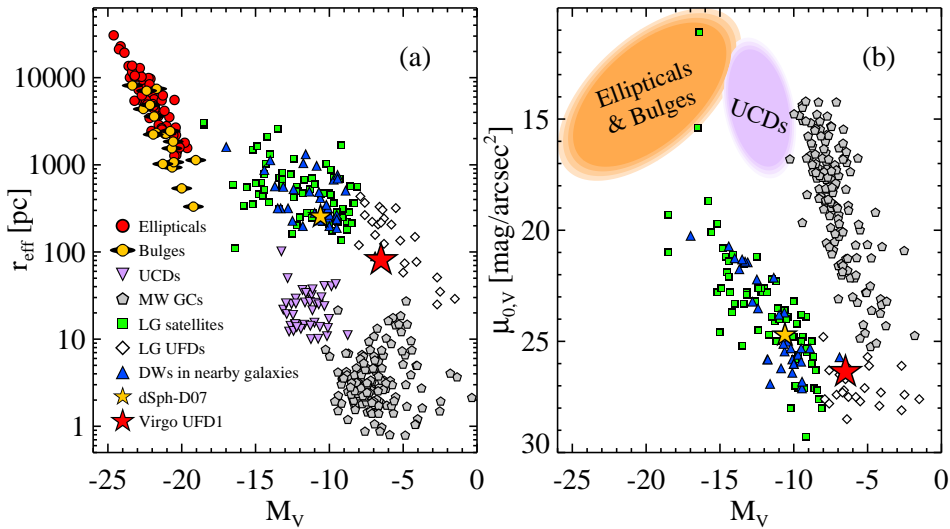} 
\caption{(a) Effective radius vs.  absolute $V$ total magnitude of the Virgo UFD1 (large starlet symbol) in comparison with those for other stellar systems.
(b) The  $V$-band central surface brightness vss  absolute $V$ total magnitude of Virgo UFD1.
Circles and lenticular symbols for the giant ellipticals and bulges in spiral galaxies, 
downward triangles for the UCDs, 
pentagons for the Milky Way globular clusters, 
squares and diamonds for the local group satellites and UFDs, 
and upward triangles for the dwarfs in M81 and M106 and the low surface brightness galaxies in M101, 
and  small starlet for Virgo dSph-D07. 
}
\label{fig_comp}
\end{figure*}

\subsection{Distance and Metallicity of Virgo UFD1}

The number of stars in the RGB of Virgo UFD1 is small, so it is not easy to derive a reliable distance to this galaxy using the TRGB method \citep{lee93}.
The $I$-band magnitude of the brightest star in the RGB is  $I=27.12\pm0.04$.
If we take the mean of the two brightest stars we obtain  $I=27.14\pm0.04$.
These stars can be considered to correspond to the TRGB.

We derived the TRGB distances to these two dwarf galaxies and intracluster stars based on our photometry.
Since they are all in the same images,
they are useful for estimating the relative distances between them.
We adopted the mean $I-$band magnitude of the two brightest stars as the TRGB for Virgo UFD1, and measured the TRGB magnitudes for Virgo dSph-D07, and intracluster stars as done in \citet{jan14}.
Derived values are : $I_{\rm TRGB} = 27.14\pm0.04$, $27.24\pm0.04$, and $26.96\pm0.02$ for Virgo UFD1, Virgo dSph-D07, and intracluster stars, respectively.
We derived the measurement error of the TRGB magnitude from 1000 simulations of bootstrap resampling.
The values for Virgo dSph-D07 and intracluster stars are consistent with those derived from the same images in previous studies: $F814W_{\rm TRGB}=27.22\pm0.15$ \citep{dur07} and $F814W_{\rm TRGB}\sim27.0$ \citep{wil07b}, respectively. 
The corresponding distance moduli, based on the TRGB calibration in \citet{riz07}, are 
$(m-M)_0 = 31.22\pm0.04$, $31.32\pm0.04$, and $31.01\pm0.02$ for Virgo UFD1, Virgo dSph-D07, and intracluster stars, respectively.
Systematic errors for these estimates are estimated to be 0.12.

We also measured distance moduli to these stellar systems based on the visual isochrone fitting on the $I - (V-I)$ CMD. 
Adopting 12 Gyr isochrones 
provided by Dartmouth group \citep{dot08}, 
we obtained $(m-M)_0 = 31.08\pm0.05$, $31.19\pm0.05$, and $30.94\pm0.04$ for Virgo UFD1, Virgo dSph-D07, and intracluster stars, respectively. 
These values are, on average, $\sim$0.1 mag smaller than, but consistent with, those from the TRGB method. 
Comparing the $V-I$ color of the red giants and the isochrones, we estimate the mean metallicity to be very low, $[$Fe/H$] = -2.4\pm0.4$.
We adopt the values from visual isochrone fitting as a distance modulus to these stellar systems.
Visual isochrone fitting determines the distance considering the stars along the isochrone, whereas, the edge detection response function in the TRGB method determined by a few stars near the TRGB level might cause a larger uncertainty in the case of Virgo UFD1.

These results show that  
 the distance modulus of Virgo dSph-D07 is $0.25\pm0.06$ (=$1.88\pm0.49$ Mpc) larger than that of intracluster stars. 
 The distance modulus of Virgo UFD1 is $0.11\pm0.07$(=$0.85\pm0.51$ Mpc) smaller than that of Virgo dSph-D7, but
$0.14\pm0.06$ (=$1.02\pm0.47$ Mpc) larger than the intracluster stars. These difference are at the level of 2$\sigma$.
These results show the following. 
First, Virgo UFD1 is indeed a member of the Virgo cluster. 
Second, Virgo UFD1 may be $\sim$1 Mpc behind the intracluster stars, but  $\sim$1 Mpc closer than Virgo dSph-D07.

\subsection{Basic Parameters of Virgo UFD1}

We derived integrated properties of this galaxy from surface photometry.
We masked out bright galaxies and bright stars that appear to be non-members of the galaxy. Then
we derived the surface photometry of the galaxy using the ELLIPSE task in IRAF.
Figure \ref{fig_surf} displays the surface brightness and integrated magnitude profiles for Virgo UFD1 as a function of mean radius ($r=\sqrt{ab}$).
We also plotted the radial density profile of detected stars, which is roughly consistent with the surface brightness profiles.
The surface brightness profiles of Virgo UFD1 are almost flat in the inner region and
decline steeply in the outer region, which is very similar to those of Virgo dSph-D07 \citep{dur07}.  
We fit the surface brightness profiles with the King model \citep{kin62} and the S\'ersic law \citep{ser68}. 
From the $V$-band King model fit, we derive the core radius, 
$r_c = 1\farcs5\pm0\farcs1$ ($120\pm8$ pc),
the tidal radius, $r_t = 3\farcs0\pm0\farcs3$ ($239\pm24$ pc),
and the concentration parameter, $c=0.30\pm0.05$.
From the $V$-band S\'ersic law fit, we obtain 
the effective radius 
$r_{\rm eff} = 1\farcs02 \pm 0\farcs09$ ($81\pm7$ pc), 
  and the central surface brightness, $\mu_{V,0} = 26.37\pm 0.05$, and
$n=0.56\pm0.06$.

The integrated magnitude profile becomes constant at
$r\gtrsim3\arcsec$.
We derived an apparent total magnitude of the galaxy,
$V = 24.66\pm0.08$  for the aperture radius $3\farcs0$. 
From this, we obtained an absolute magnitude, $M_V = -6.5\pm0.2$,
whose error includes the errors due to photometry and distance measurement.
We also estimated total magnitude of the galaxy, integrating the luminosity of the resolved red giants 
and unresolved stars assuming luminosity function with a power-law index $\alpha=0.3$,
obtaining $V \approx 25.0$. 
This value is similar to the value derived from integrated photometry.
Thus, Virgo UFD1 is much smaller and fainter than the classical dSphs in the Local Group and smaller and fainter than any known Virgo galaxies.
Table 1 lists the basic parameters of Virgo UFD1 derived in this study.

\begin{deluxetable}{lc}
\tabletypesize{\footnotesize}
\setlength{\tabcolsep}{0.05in}
\tablecaption{Basic Parameters of Virgo UFD1}
\tablewidth{0pt}
\tablehead{ \colhead{Parameter} & \colhead{Value$^a$} }
\startdata
R.A.(2000) 	& $12^h28^m06.^s061$			 	\\
Dec(2000) 	& $12\arcdeg33\arcmin47\farcs61$ 	\\
Type		& UFD							 	\\
Distance, $(m-M)_0$ & $31.08\pm0.05$ ($16.4\pm0.4$ Mpc)  	\\
Total magnitude, $V^T$	& $24.66\pm0.08$   \\
Total color, $V^T-I^T$	& $1.06\pm0.10$  \\
Ellipticity, $e=(a-b)/a$& $0.1\pm0.1$   \\
Absolute magnitude, $M_V$	& $-6.5\pm0.2$  \\
Position angle			& $130\arcdeg\pm10\arcdeg$   \\
Core radius ($r_{\rm core}$), $V, I$	& $1\farcs5\pm0\farcs1$, $1\farcs5\pm0\farcs1$     \\
Tidal radius ($r_{\rm tidal}$), $V, I$	& $3\farcs0\pm0\farcs3$, $3\farcs1\pm0\farcs3$    \\
Sersic Index (n), $V, I$ 		& $0.56\pm0.06$, $0.52\pm0.05$    \\
Effective radius ($r_{\rm eff}$), $V, I$	& $1\farcs02\pm0\farcs09$, $1\farcs01\pm0\farcs09$    \\
Central surface brightness	& $26.37\pm0.05$, $25.34\pm0.04$     \\
\hline
\tablenotetext{a}{Derived in this study  }
\enddata

\label{tab_param}
\end{deluxetable}
\section{Discussion and Conclusion}

\subsection{Comparison with Other Dwarf Galaxies}

In Figure 4 we compare the structural parameters of Virgo UFD1 and those for
other dwarf galaxies in the nearby universe:
Local Group dwarfs including dSphs and UFDs \citep{mcc12,san12,tol13},
dwarf galaxies in nearby galaxies (M81 group \citep{chi13} and M106 \citep{kim11}), low surface brightness galaxies in  M101 \citep{mer14},
and Virgo dSph-D07 \citep{dur07}.
We also plotted the globular clusters in the Milky Way galaxy \citep{har10},
ultra-compact dwarf objects (UCDs) in nearby galaxies \citep{bro11, nor11, pen14},
and elliptical galaxies \citep{ben93} for reference.
  
The half-light radius and absolute magnitude of Virgo UFD1 follow the relation for other UFDs in the Local Group extending out to that for dSphs, but they are separated from those for globular clusters and UCDs. Virgo UFD1 is larger, but fainter, than UCDs.
It is as bright as a globular cluster, but is significantly larger than the latter.
The central surface brightness and  absolute magnitude of this new galaxy are also similar to those of the Local Group UFDs.  
It turns out that the properties of Virgo UFD1 are similar to those of Local Group UFDs.

\subsection{Fossils of the First Galaxies?}

According to the current LCDM model, the minimum mass of the dark matter subhalos that can form before reionization is lower than $10^8-10^9  M_\odot$. The first galaxies formed from these subhalos formed most of their stars before reionization ($z>6$, $>12$ Gyr ago). If these stars survive in the local universe today, they can be observed as fossils of the first galaxies \citep{bul01,ric05,ric10,bov11a,bov11b}. 
Some UFDs in the Local Group are strong candidates for these fossils \citep{bro13,wei14}.

It is not possible to derive a detailed star formation history of Virgo UFD1, because our photometry does not reach the  main-sequence turnoff. 
However, even the current photometry provides a useful clue for tracing the early history of this galaxy.
Virgo UFD1 shows only metal-poor RGB stars and no AGB stars, while  Virgo dSph-D07
shows a small number of AGB star candidates above the TRGB.  
This indicates that Virgo UFD1 is very old (older than 10 Gyr). 
It may be as old as the six Local Group UFDs that show only stars older than 12 Gyr \citep{bro13}.  

It is located close to the Virgo center,  far (more than 180 kpc) from any nearby massive galaxies.
The projected distance between this new galaxy and M87 is large, but smaller than the viral radius, $R_V\simeq470$ kpc \citep{doh09}.
We cannot tell the spatial separation between this and other massive galaxies
at the moment, but it must be larger than the projected distance.
 \citet{bir10} derived a TRGB distance to M87 from the $HST$ photometry of an inner field,
$(m-M)_0 = 31.12\pm0.14$ ($16.7\pm0.9$ Mpc).
This value is similar, within errors, to that to Virgo UFD1. Therefore we cannot tell whether Virgo UFD1 is located within the virial radius of M87 or other massive galaxies.

We checked any hint of tidal interaction (tidal shredding) from the morphology and structure of this new galaxy in the images, finding no clear signature of tidal interaction around this galaxy. 
\citet{dur07} also found that there is no hint of tidal feature around Virgo dSph-D07.
\citet{dur07} discussed a possibility that dSph-D07 may be behind the Virgo center,
at a similar distance to M86  and that dSph-D07 may be a member of the M86 group and may as if recently be infalling to the Virgo Center recently.
Considering its old age (based on narrow RGB with no AGB stars), low metallicity, and spatial location in the intracluster field, 
Virgo UFD1 may be a fossil remnant of the first galaxies. 

\subsection{Missing Satellite Dwarfs}

The number of dark satellite subhalos around massive galaxies like the Milky Way Galaxy predicted from simulations based on the hierarchical formation scenario, including cold dark matter, was of the order of magnitude larger than
the number of known satellite dwarf galaxies around the Milky Way Galaxy and M31 \citep{kly99,moo99}. It is called a massing satellite dwarf problem. 
Since the introduction of this problem, much progress has been made
in both observations and theories to solve this problem, but it is still a hot issue \citep{kra10,bel13}.
From the observational point of view, UFDs are a strong candidate for understanding this problem.
The number of known UFDs in the Local Group is $\sim20$ \citep{bel13}.
This number is still increasing and will continue to do so with the advent of new wide-field surveys \citep{bel14,lae14}.

The discovery of one UFD in Virgo indicates that UFDs are not unique to the Local Group, but may be common throughout in the universe. There may be thousands of UFDs in Virgo, and hundreds of UFDs around each massive galaxy.  Some true fossils of the first galaxies may be wandering in the intracluster field. Thus, Virgo may provide another laboratory in which to investigate the missing
satellite problem. We need a deep survey with high spatial resolution (like the $HST$) to find more UFDs in Virgo. 
 
This work was supported by the National Research Foundation of Korea (NRF) grant
funded by the Korea Government (MSIP) (No. 2013R1A2A2A05005120).
This paper is based on image data obtained from the Multimission Archive at the Space Telescope Science Institute (MAST).




\end{document}